\documentclass[prl,twocolumn,amsmath,nofootinbib,amssymb,superscriptaddress]{revtex4}
\usepackage{graphicx}

\usepackage{epsfig,psfrag,amsmath,amssymb,float}
\begin{document}

\newcommand{\ltwid}{\mathrel{\raise.3ex\hbox{$<$\kern-.75em\lower1ex\hbox{$\sim$}}}}
\newcommand{\gtwid}{\mathrel{\raise.3ex\hbox{$>$\kern-.75em\lower1ex\hbox{$\sim$}}}}
\def\K{{\bf{K}}}
\def\Q{{\bf{Q}}}
\def\Gbar{\bar{G}}
\def\tk{\tilde{\bf{k}}}
\def\k{{\bf{k}}}

\title{The Structure of the Pairing Interaction 
 in the 2D Hubbard Model}

\author{T.A.~Maier}
\affiliation{Computer Science and Mathematics Division,\\
 Oak Ridge National Laboratory, 
Oak Ridge, TN 37831-6164}
\email{maierta@ornl.gov}

\author{M.S.~Jarrell}
\affiliation{Department of Physics,\\
 University of Cincinnati, Cincinnati, OH 45221}
\email{jarrell@physics.uc.edu}

\author{D.J.~Scalapino}
\affiliation{Department of Physics,\\
 University of California, Santa Barbara, CA 93106-9530}
\email{djs@vulcan2.physics.ucsb.edu}

\date{\today}

\begin{abstract}

  Dynamic cluster Monte Carlo calculations for the doped
  two-dimensional Hubbard model are used to study the irreducible
  particle-particle vertex responsible for $d_{x^2-y^2}$ pairing in
  this model.  This vertex increases with increasing momentum transfer
  and decreases when the energy transfer exceeds a scale associated
  with the $Q=(\pi, \pi)$ spin susceptibility. Using an exact
  decomposition of this vertex into a fully irreducible two-fermion
  vertex and charge and magnetic exchange channels, the dominant part
  of the effective pairing interaction is found to come from the
  magnetic, spin $S=1$ exchange channel.

\end{abstract}

\maketitle

%==========BODY OF PAPER =========================================

%\section*{} Shortly after the discovery of high-temperature superconductors,
%the Hubbard and related models were identified to play a central role in their
%description.
Numerical studies of the doped two-dimensional Hubbard model have shown that
strong $d_{x^2-y^2}$ pairing correlations develop as the temperature is lowered
\cite{BSW93} and recent work has provided evidence that the Hubbard model near
half filling does have a superconducting ground state \cite{Mai05,Sen05,Kan05}.
Inspite of this progress, the goal of using numerical methods to determine the
nature of the pairing mechanism has proven elusive.  In this paper we develop a
new approach which combines numerical and diagrammatic methods to determine the
structure of the pairing interaction in the doped two-dimensional Hubbard
model.  Our study will focus on the 4-point vertex calculated with a Quantum
Monte Carlo dynamic cluster approximation\cite{jarrell:qmc, hettler:dca,
maier:rev}.  From this vertex and the QMC/DCA results for the single particle
Green's function, we have determined the irreducible particle-particle and
particle-hole vertices. The leading low temperature eigenvalue of the
Bethe-Salpeter equation for the particle-particle channel is shown to have
$d_{x^2-y^2}$ symmetry.  We then examine the momentum and energy dependence of
the irreducible particle-particle vertex.  Decomposing this vertex into the sum
of a fully irreducible two-fermion vertex and particle-hole exchange magnetic
($S=1$) and charge density ($S=0$) channels, we find that the dominant
contribution to the pairing interaction comes from the magnetic ($S=1$)
exchange.

The Hubbard model that we study has a near-neighbor, one-electron hopping $t$
and an on-site Coulomb interaction $U$.
\begin{eqnarray}
  H&=&-t \sum_{\langle ij\rangle\sigma}\ \left(c^\dagger_{i\sigma} c_{j\sigma} +
    c^\dagger_{j\sigma} c_{i\sigma}\right) + U \sum_i n_{i\uparrow} n_{i\downarrow}\nonumber\\
  & &- \mu \sum_{i\sigma} c^\dagger_{i\sigma} c_{i\sigma}\, .
\label{one}
\end{eqnarray}
We will take $U/t=4$ and adjust $\mu$ so that the average site occupancy
$\langle n_i\rangle =0.85$.  We have carried out dynamical cluster Monte Carlo
calculations \cite{jarrell:qmc, hettler:dca, maier:rev} for the 24-site
$k$-cluster shown in the inset of Fig.~2.

For a $2D$ system the dynamical cluster approximation maps the original lattice
model onto a periodic cluster of size $N_c=L_c^2$ embedded in a self-consistent
host \cite {jarrell:qmc, hettler:dca,maier:rev}. The essential assumption is
that short-range quantities, such as the self energy and its functional
derivatives (the irreducible vertex functions) are well represented as diagrams
constructed from the coarse-grained Green's function.  To this end, the first
Brillouin zone is divided into $N_c$ cells, with each cell represented by its
center wave-vector $\K$ surrounded by $N/N_c$ lattice wavevectors labeled by
$\tk$. The reduction of the $N$-site lattice problem to an effective $N_c$ site
cluster problem is achieved by coarse-graining the single-particle Green's
function, {\it i.e.} averaging $G(\K+\tilde{\k})$ over the $\tk$ within a cell
which converges to a cluster Green's function $G_c(\K)$. Consequently, the
compact Feynman diagrams constructed from $G_c(\K)$ collapse onto those of an
effective cluster problem embedded in a host which accounts for the
fluctuations arising from the hopping of electrons between the cluster and the
rest of the system.  The compact cluster quantities are then used to calculate
the corresponding lattice quantities.
  
\begin{figure}[htb]
\begin{center}
\includegraphics*[width=3.3in]{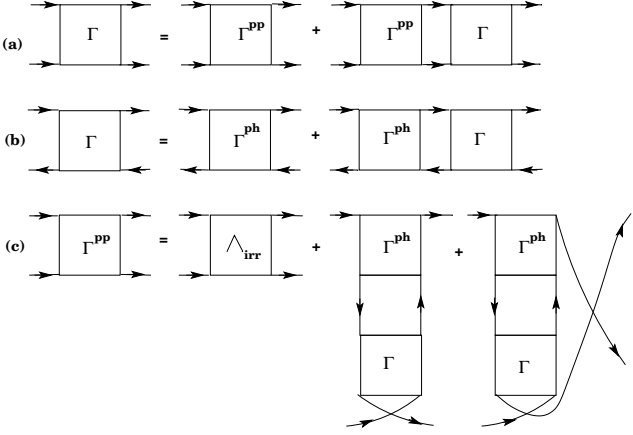} 
\caption{Bethe-Salpeter equations for (a) the particle-particle and
  (b) the particle-hole channels showing the relationship between the
  full vertex, the particle-particle irreducible vertex $\Gamma^{\rm
    pp}$, and the particle-hole irreducible vertex $\Gamma^{\rm ph}$,
  respectively. (c) Decomposition of the irreducible particle-particle
  vertex $\Gamma^{\rm pp}$ into a fully irreducible two-fermion vertex
  $\wedge_{\rm irr}$ plus contributions from the particle-hole
  channels. All diagrams represent DCA cluster quantities, including
  the Green function legs.}
\label{fig:2pdiagrams}
\end{center}
\end{figure}

For example, the DCA cluster one- and two-particle Green's functions
that we calculate have the standard finite temperature definitions
\begin{subequations}
\label{two}
\begin{equation}
  G_{c\sigma} (X_2, X_1) = 
  - \left\langle T_\tau c_\sigma (X_2)\, c^\dagger_\sigma (X_1)\right\rangle
\label{two a}
\end{equation}
and
\begin{eqnarray}
  G_{c 2\sigma_4\cdots\sigma_1} (X_4, X_3; X_2, X_1) = \hspace*{-4cm}&&\nonumber\\
  &-& \left\langle T_\tau c_{\sigma_4} (X_4)
    \, c_{\sigma_3} (X_3)\, c^\dagger_{\sigma_2} (X_2)\, c^\dagger_{\sigma_1} (X_1)\right\rangle
  \, .
\label {two b}
\end{eqnarray}
\end{subequations}
Here, $X_\ell = ({\bf X}_{\ell}, \tau_\ell)$, where ${\bf X}_{\ell}$
denotes a site in the DCA cluster, $\tau_{\ell}$ the imaginary time,
$T_\tau$ is the usual $\tau$-ordering operator, and
$c^{(\dagger)}_\sigma (X_2)$ creates (destroys) a particle on the
cluster with spin $\sigma$.  Fourier transforming on both the cluster
space and imaginary time variables gives $G_c (K)$ and $G_{c2}(K_4,
K_3; K_2, K_1)$ with $K=({\K}, i\omega_n, \sigma)$.  Using $G_c(K)$
and $G_{c2}(K_4, K_3; K_2, K_1)$, one can extract the cluster
four-point vertex $\Gamma$ from
\begin{eqnarray}
\label{three}
G_{c2} (K_4, K_3; K_2, K_1) = \hspace*{-3cm} &&\nonumber\\
&-& G_c(K_1)\, G_c(K_2)\, \left[\delta_{K_1, K_4} \delta_{K_2, K_3}
  - \delta_{K_1, K_3} \delta_{K_2, K_4}\right]\nonumber \\
&+& \frac{T}{N}\ \delta_{K_1+K_2, K_3+K_4} G_c(K_4)\, G_c(K_3) \Gamma\, (K_4, K_3; K_2, K_1)\nonumber \\
&\times&  G_c(K_2)\, G_c(K_1)\, .
\end{eqnarray}
Then, using $G_c$ and $\Gamma$, one can determine the irreducible
particle-particle and particle-hole vertices $\Gamma^{\rm pp}$ and
$\Gamma^{\rm ph}$ from the Bethe-Salpeter equations shown in
Fig.~\ref{fig:2pdiagrams}(a) and (b).  There is a second particle-hole
vertex but it is simply related to $\Gamma^{\rm ph}$.  Note that
$\Gamma^{\rm ph}$ and $\Gamma^{\rm pp}$ don't have a c suffix, since
both the lattice and the cluster share these compact quantities.
Because of the rotational invariance of the Hubbard model, it is
convenient to separate the particle-particle channels into singlet and
triplet and the particle-hole channels into a magnetic part which
carries spin $S=1$ and a charge density part which has $S=0$.

In order to determine the nature of the low temperature correlations,
we use these irreducible vertices and the lattice single particle
Green's function to calculate the Bethe-Salpeter eigenvalues and
eigenfunctions in various channels. For example, in the
particle-particle channel
\begin{eqnarray}
  -\frac{T}{N}\ \sum_{k^\prime} \Gamma^{\rm pp} \left(K, -K; K^\prime, -K^\prime\right)\hspace*{-4cm}&&\nonumber\\
  &\times& G_\uparrow (k^\prime)\, G_\downarrow (-k^\prime)\, 
  \phi_\alpha (K^\prime) =
  \lambda_\alpha \phi_\alpha (K)
\label{four}
\end{eqnarray}
with a similar equation using $\Gamma^{\rm ph}$ for the particle-hole
channel.  Here, the sum over $k^\prime$ denotes a sum over both
momentum {\boldmath$k$}$^\prime$ and Matsubara $\omega_{n^\prime}$
variables. We decompose $\k^\prime = \K^\prime + \tk^\prime $.  By
assumption, irreducible quantities like $\Gamma^{pp}$ and
$\phi_\alpha$ do not depend on $\tk^\prime$, allowing us to
coarse-grain the Green function legs, yielding an equation that
depends only on coarse-grained and cluster \ quantities
\begin{eqnarray}
  -\frac{T}{N_c}\ \sum_{K^\prime} \Gamma^{\rm pp} \left(K, -K; K^\prime, -K^\prime\right)\,
  {{\bar{\chi}}_0^{\rm pp}}(K')
  \, \phi_\alpha (K^\prime) = &&\nonumber\\
  && \hspace{-2cm}\lambda_\alpha \phi_\alpha (K)
\label{Eq:eigvalcg}
\end{eqnarray}
with ${{\bar{\chi}}_0^{\rm pp}}(K') = \frac{N_c}{N}
\sum_{\tk^{\prime}} G_\uparrow (\K^\prime+\tk^\prime) \, G_\downarrow
(-\K^\prime-\tk^\prime)$.

\begin{figure}[htb]
\begin{center}
\includegraphics*[width=3.5in]{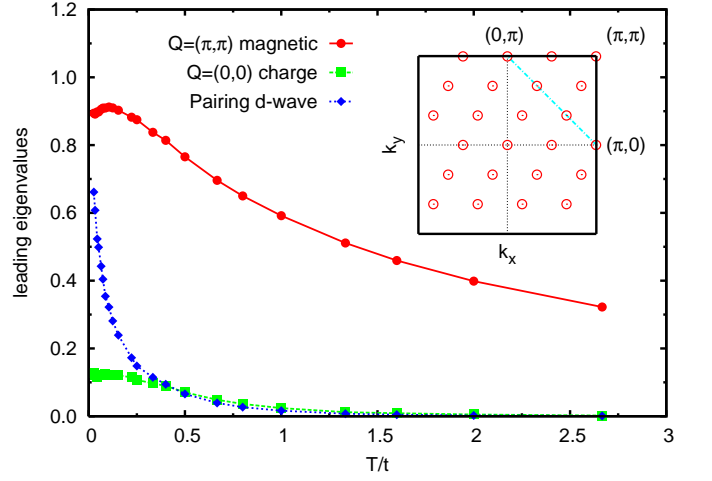}
\caption{Leading eigenvalues of the Bethe-Salpeter equation (e.g.\
  Eq.~\ref{Eq:eigvalcg}) in various channels for $U/t=4$ and a site
  occupation $\langle n\rangle=0.85$. The ${\bf Q}=(\pi, \pi)$,
  $\omega_m=0$, $S=1$ magnetic eigenvalue is seen to saturate at low
  temperatures.  The leading eigenvalue in the singlet ${\bf
    Q}=(0,0)$, $\omega_m=0$ particle-particle channel has
  $d_{x^2-y^2}$ symmetry and increases toward 1 at low temperatures
  \cite{Mai05}.  The largest charge density eigenvalue occurs in the
  ${\bf Q}=(0, 0)$, $\omega_m=0$ channel and saturates at a small
  value. The inset shows the distribution of $k$-points for the
  24-site cluster we have studied.}
\label{fig:eigs}
\end{center}
\end{figure}

In Fig.~\ref{fig:eigs} we show the leading eigenvalue versus temperature for
the pairing, charge density, and magnetic channels for $U/t=4$ and $\langle
n\rangle=0.85$. As the temperature is reduced, the leading particle-hole
eigenvalue occurs in the magnetic channel and has a center of mass momentum
${\bf Q}=(\pi, \pi)$ and $\omega_m=0$. Previous Monte Carlo calculations on
8$\times$8 lattices show that for this doping the peak response is slightly
shifted from $(\pi,\pi)$, but our 24-site cluster lacks the resolution to see
this \cite{Mor90}.  This antiferromagnetic eigenvalue grows and then saturates
at low temperatures. The leading particle-particle eigenvalue is a spin singlet
and, as shown in the inset of Fig.~3, its eigenfunction $\phi_{d_{x^2-y^2}}$
has $d_{x^2-y^2}$ symmetry.  The $\omega_n$ frequency dependence of the
normalized gap function $\phi_{d_{x^2-y^2}}(\K, \omega_n)$ at the antinodal
point $\K=(\pi,0)$ is plotted in Fig.~3. As shown, it is even in $\omega_n$,
corresponding to a $d_{x^2-y^2}$-wave singlet, even frequency pairing.  Also
plotted in this figure is the $\omega_m$-dependence of the ${\bf Q}=(\pi, \pi)$
spin susceptibility $\chi({\bf Q}, \omega_m)$ normalized to coincide with
$\phi_{d_{x^2-y^2}}(\K, \omega_n)$ at $\omega_n=\pi T$. The Matsubara frequency
which enters the gap function corresponds to a fermion frequency
$\omega_n=(2n+1) \pi T$, while $\omega_m=2m\pi T$ for the spin susceptibility,
leading to the interlacing of points shown in Fig.~3. From the momentum and
frequency dependence of the gap function $\phi(\K, \omega_n)$, it follows that
the irreducible particle-particle vertex is an increasing function of the
momentum transfer and is characterized by the same energy scale that enters the
spin susceptibility $\chi ({\bf Q},\omega_m)$.  At larger values of $U$ it will
be interesting to see if there is an increased tendency for a finite response
at large Matsubara frequencies indicating a contribution from the upper Hubbard
band.

\begin{figure}[h] \begin{center}
	\includegraphics[width=3.5in]{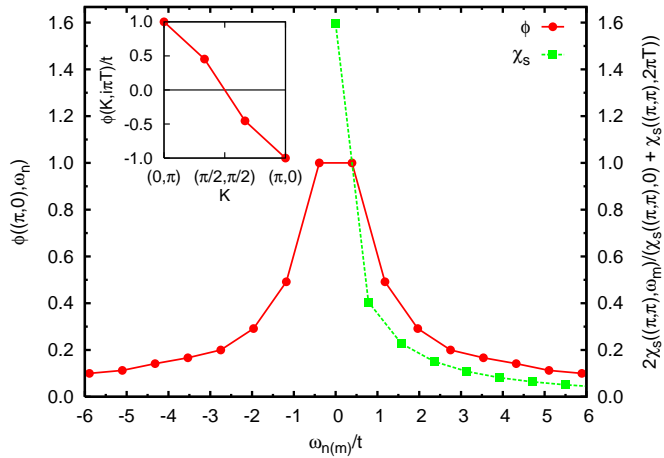} \caption{The
	Matsubara frequency dependence of the eigenfunction $\phi_{d_{x^2-y^2}}
	(\K, \omega_n)$ of the leading particle-particle eigenvalue of Fig.~2
	for ${\bf K}=(\pi, 0)$ normalized to $\phi(\K,\pi T)$ (red, solid).
	Here, $\omega_n=(2n+1)\pi T$ with $T=0.125t$. The Matsubara frequency
	dependence of the normalized magnetic spin susceptibility $2\chi({\bf
	Q}, \omega_m)/[\chi({\bf Q},0)+\chi({\bf Q},2\pi T)]$ for ${\bf
	Q}=(\pi, \pi)$ versus $\omega_m=2m\pi T$ (green, dashed). Inset: The
	momentum dependence of the eigenfunction $\phi_{d_{x^2-y^2}} (\K, \pi
	T)$ normalized to $\phi_{d_{x^2-y^2}} \left( (0,\pi),\pi T \right)$
	shows its $d_{x^2-y^2}$ symmetry. Here, $\omega_n=\pi T$ and the
	momentum values correspond to values of $\K$ which lay along the dashed
	line shown in the inset of Fig.~2.} \end{center} \end{figure}

To learn more about the mechanism responsible for $d_{x^2-y^2}$ pairing in the
doped Hubbard model, it is useful to decompose the pairing interaction
$\Gamma^{\rm pp}$ as shown in Fig.~1c. Here, the irreducible particle-particle
vertex is given as a combination of a fully irreducible two-fermion vertex
$\wedge_{\rm irr}$ and partially reducible particle-hole exchange contributions
\cite{PW89, EB98}. For the even frequency, even momentum part of the
irreducible particle-particle vertex $\Gamma^{\rm pp}_{\rm
even}(K,K')=1/2[\Gamma^{\rm pp}(K,K')+\Gamma^{\rm pp}(K,-K')]$, we obtain
\begin{eqnarray}
	\Gamma^{\rm pp}_{\rm even} (K,K') 
  &=&  \wedge_{\rm irr} (K, K')\nonumber\\
  &+&  \frac{1}{2} \Phi_d(K,K') + \frac{3}{2} \Phi_m (K,K')  
\label{five}
\end{eqnarray}
with $K=(\K,i\omega_n)$. The subscripts $d$ and $m$ denote the charge density $(S=0)$ and
magnetic $(S=1)$ particle-hole channels
\begin{eqnarray}
	\label{six}
	\Phi_{d/m}(K,K') = \hspace*{-2.5cm}&&\\
	&&\frac{1}{2}\left[ \Gamma_{d/m}(K-K';K',-K)-\Gamma_{d/m}^{\rm ph}(K-K';K',-K)\right.\nonumber\\
	&+& \left. \Gamma_{d/m}(K+K';-K',-K)-\Gamma_{d/m}^{\rm ph}(K+K';-K',-K)\right]\,. \nonumber
\end{eqnarray} 
The center of mass and relative wave vectors and frequencies in these channels
are labeled by the first, second and third arguments respectively.  Using the
Monte Carlo results for $G$ and $\Gamma$, we have solved the $t$-matrix
equations shown in Figs. 1(a) and (b) to determine $\Gamma^{\rm pp}$, $\Phi_d$
and $\Phi_m$. Then, substituting these into Eq.~\eqref{five}, we have determined
the fully irreducible vertex $\wedge_{\rm irr}$.

\begin{figure*}[htb]
\begin{center}
\includegraphics[width=5.5in]{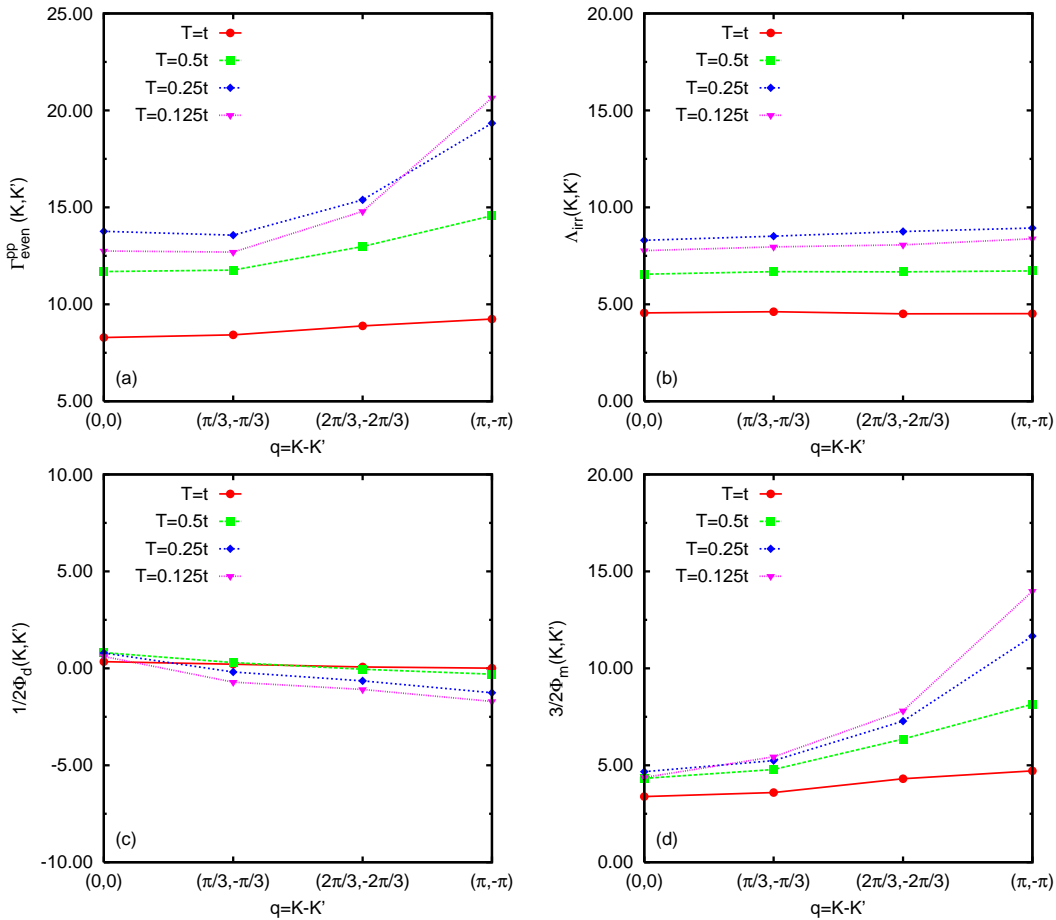}
\caption{(a) The irreducible particle-particle vertex $\Gamma^{\rm pp}$ versus
${\bf q}=\K-\K^\prime$ for various temperatures with
$\omega_n=\omega_{n^\prime}=\pi T$. Here, $\K=(\pi, 0)$ and $\K^\prime$ moves
along the momentum values of the 24-site cluster which lay on the dashed line
shown in the inset of Fig.~2.  Note that the interaction increases with the
momentum transfer as expected for a d-wave pairing interaction.  (b) The ${\bf
q}$-dependence of the fully irreducible two-fermion vertex $\wedge_{\rm irr}$.
(c) The ${\bf q}$-dependence of the charge density $(S=0)$ channel
$\frac{1}{2}\Phi_d$ for the same set
of temperatures.  (d) The ${\bf q}$-dependence of the magnetic $(S=1)$ channel
$\frac{3}{2} \Phi_m$.}
\end{center}
\end{figure*}

Monte Carlo results for the irreducible particle-particle vertex $\Gamma^{\rm
pp}$ obtained from the 24-site cluster approximation are shown in Fig.~4a.
Here, we set $\omega_n=\omega_n'=\pi T$, $\K=(\pi, 0)$ and $\K^\prime$ takes
momentum values along the dashed line shown in the inset of Fig.~2. 
%For these values of the momentum, ${\bf K} + {\bf K}^\prime$ can be replaced
%by ${\bf K} - {\bf K}^\prime$ in the last term. 
As the temperature is lowered, $\Gamma^{\rm pp}$ increases as the momentum
transfer ${\bf q}={\bf K} - {\bf K}^\prime$ increases as one expects for a
d-wave pairing interaction.  To understand the origin of this behavior, the
contributions of the particle-hole $(S=0)$ charge density and $(S=1)$ magnetic
channels are plotted in Figs.~4c and 4d respectively and the contribution from
the fully irreducible vertex $\wedge_{\rm irr}$ is shown in Fig.~4b.  It is
clear that the dominant contribution to $\Gamma^{\rm pp}$ comes from the $S=1$
magnetic channel.  The charge density channel and the fully irreducible vertex
are basically flat in momentum and change relatively little as the temperature
is reduced.  Thus, based upon the decomposition of the irreducible
particle-particle interaction shown in Fig.~4, we conclude that the pairing
mechanism in the doped two-dimensional Hubbard model is mediated by the
exchange of $S=1$ particle-hole spin-fluctuations.

To summarize, we have studied the pairing interaction $\Gamma^{\rm pp}$ of a
doped $\langle n \rangle=0.85$, two-dimensional Hubbard model with $U/t=4$.  We
found that the eigenfunction $\phi(\K,i\omega_n)$ of the leading low
temperature eigenvalue in the particle-particle pairing channel is an even
frequency singlet with $d_{x^2-y^2}$ symmetry. The momentum and frequency
dependence of $\phi(\K,i\omega_n)$ imply that $\Gamma^{\rm pp}$ increases as
the momentum transfer ${\bf q}=\K-\K^\prime$ increases and that its dynamics is
set by the same characteristic energy scale as the spin susceptibility.  It was
also found to increase as the temperature was lowered, saturating when the
leading antiferromagnetic eigenvalue stopped growing.  Finally, using an exact
decomposition of $\Gamma^{\rm pp}$, we showed that the dominant contribution to
this interaction comes from the $S=1$ particle-hole channel.  We believe that
the calculation and analysis of the four-point vertex provides a useful,
unbiased method for determining the nature of the leading correlations of
interacting many-electron systems and the structure of the mechanisms
responsible for them.

\acknowledgments We acknowledge useful discussions with W.~Putikka, R.~Sugar,
and S.-C.~Zhang.  This research was enabled by computational resources of the
Center for Computational Sciences at Oak Ridge National Laboratory and was
conducted at the Center for Nanophase Materials Sciences, which is funded by
the Division of Scientific User Facilities, U.S. Department of Energy.  This
research was supported by NSF Grants DMR-0312680 and DMR02-11166.


\begin{thebibliography}{99}

\bibitem{BSW93} N.~Bulut, D.J.~Scalapino, and S.R.~White, {\sl Phys.~Rev.~B}
	{\bf 47} R6157 (1993); {\sl ibid.} {\bf 47}, 14599 (1993).

\bibitem{Mai05} T.A.~Maier, M.~Jarrell, T.C.~Schulthess, P.R.C.~Kent, and
	J.B.~White, Phys. Rev. Lett. {\bf 95}, 237001 (2005).

\bibitem{Sen05} David S\'en\'echal, P.-L. Lavertu, M.-A. Marois, and A.-M. S.
	Tremblay, Phys. Rev. Lett. {\bf 94}, 156404 (2005).

\bibitem{Kan05} S. S. Kancharla, M. Civelli, M. Capone, B. Kyung, D.
	S\'en\'echal, G. Kotliar, A.-M.S. Tremblay, preprint cond-mat/0508205
	(2005).

\bibitem{jarrell:qmc} M.~Jarrell, T. Maier, C. Huscroft and S. Moukouri, {\sl
	Phys.~Rev.~B} {\bf 64}, 195130 (2001).


\bibitem{hettler:dca} M.H.~Hettler, A.N. Tahvildar-Zadeh, M. Jarrell, T.
	Pruschke, and H.R. Krishnamurthy, {\sl Phys.~Rev.~B} {\bf 58}, R7475
	(1998); M.H.~Hettler, M. Mukherjee, M. Jarrell, and H.R. Krishnamurthy,
	{\sl ibid.} {\bf 61}, 12739 (2000).

\bibitem{maier:rev} Th.~Maier {\em et al.}, Rev. Mod. Phys. {\bf 77}, 1027
	(2005).

\bibitem{Mor90} A.~Moreo, D.J. Scalapino, R.L. Sugar, S.R. White, and N.E.
	Bickers, {\sl Phys.~Rev.~B} {\bf 41}, 2313 (1990).

\bibitem{PW89} M.~Pfitzner and P.~W\"olfle, {\sl Phys.~Rev.~B} {\bf 35}, 4699
	(1989).  \bibitem{EB98} G.~Esirgen and N.E.~Bickers, {\sl
	Phys.~Rev.~B} {\bf 57}, 5376 (1998).


\end{thebibliography}
\end{document}